\documentclass[traditabstract]{aa}

\usepackage{amssymb}
\usepackage{natbib}
\usepackage{txfonts}
\usepackage{graphicx}

\def\mstar{M_*}
\def\mrecent{M_{recent}}

\begin{document}

\title{Evidence for short-lived SN Ia progenitors}
\titlerunning{Evidence for short-lived SN Ia progenitors}
\authorrunning{Aubourg et al.}

\author{\'Eric
  Aubourg\inst{1,2,3}, Rita Tojeiro\inst{4},
  Raul Jimenez\inst{5}, Alan Heavens\inst{4},
   Michael A. Strauss\inst{3} \and David N. Spergel\inst{3}}

\institute{Astroparticule et Cosmologie APC, UMR 7164, Universit\'e Paris Diderot,
10 rue Alice Domon et L\'eonie Duquet,
75205 Paris cedex 13, France
\and
CEA, Irfu, SPP, Centre de Saclay, 91191 Gif sur Yvette cedex, France
\and
Department of Astrophysical Sciences, Peyton Hall, Princeton University, Princeton, NJ 08544, USA
\and
SUPA, Institute for Astronomy, University of Edinburgh, Royal Observatory, Blackford Hill, Edinburgh EH9-3HJ, UK
\and
Institute of Space Sciences (CSIC-IEEC)/ICREA, Campus UAB, Barcelona 08193, Spain}

\abstract{We use the VESPA algorithm and spectra from the Sloan Digital Sky
Survey to investigate the star formation history of the host galaxies
of 257 Type Ia supernovae.  We find 5$\sigma$ evidence for a
short-lived population of progenitors with lifetimes of less than 180 Myr, indicating a Type Ia supernova channel arising from stars in the
mass range $\sim$3.5-8 $M_\odot$.  As standardizeable candles,
Type Ia supernovae play an important 
role in determining the expansion history of the Universe, but to be
useful for future cosmological surveys, the peak luminosity needs to
be free of uncorrected systematic effects at the level of 1-2\%.  If the different progenitor routes lead to supernovae with even moderately small differences in properties, then these need to be corrected for separately, or they could lead to a systematic bias in future supernovae surveys, as the prompt route is likely to increase in importance at high redshift.  VESPA analysis of hosts could be a valuable tool in this, by identifying which progenitor route is most likely.}


\maketitle

\section{Introduction}

The relationship between apparent peak brightness and redshift of Type Ia supernovae
(SNe Ia) depends on the cosmological model; this has provided the most
direct evidence for the accelerated expansion of the Universe \citep{riess98,perl99}. Current SN Ia surveys such as SNLS \citep{snls06}, ESSENCE \citep{essence}, and GOODS-SN \citep{snhst} are contributing to current constraints on
cosmological parameters, and SNe Ia will continue to be important for cosmological
constraints in the next generation of surveys such as the JDEM candidates
ADEPT, DESTINY \citep{destiny} and SNAP \citep{snap}.

Type Ia supernovae are interpreted as the thermonuclear explosion of a white
dwarf that has reached the Chandrasekhar mass and thus has become
unstable, probably through accretion from a companion star
(the single-degenerate scenario) or merging with another
white dwarf (the double-degenerate scenario). However, no fully consistent
model of a SN Ia explosion has yet been built.

The natural scatter in SNe Ia peak luminosities covers roughly one
magnitude; the rms peak luminosity is 0.45 mag after excluding
outliers.
Empirical correlations based on light curve shape \citep{phil93} or
intrinsic color \citep{vdb, tripp}
allow reduction of the intrinsic scatter to about 0.13 mag  \citep{snls06}, making them
usable for cosmological measurements.

However, it is not yet known how much of the residual scatter is
correlated with physical parameters that could evolve with redshift,
and thus bias the measurement of cosmological parameters.
In order for SNe to be useful for
constraining dark energy at the level expected in future SN satellite
experiments, the evolution of
luminosity at a given light-curve shape over the probed redshift range
must be less than 1-2\% \citep{how07, sarkar08}.

There are several observational indications for a range of delay times
between the birth of the progenitor system and the explosion of
the SN Ia, leading some authors to envision the existence of at least
two different populations of SN Ia populating slightly different regions of
stretch-brightness parameter space (Sullivan et al. 2006). It is not yet known if they
are described (at the percent level) by the same Phillips relation; if they correspond
to different production channels there is no reason for the Phillips relations to be identical.
In any case, the SN population may not be a one-dimensional family,
but may depend on other parameters (at an as-yet unknown level) such
as metallicity and delay time. 
This would result in a source of scatter in the SN Ia Hubble diagram that could be reduced 
with measurements of those additional parameters. 
Moreover, if the Phillips relations are different and the
relative numbers of SNe in the two putative
populations evolve with redshift, the values of
cosmological parameters derived from the Hubble relation will be
biased if we do not determine population-dependent corrections.
More generally, any extra parameter (age, metallicity, production channel...) in the Phillips relations
that is correlated with redshift will bias cosmological measurements. It is therefore essential
to assess the size of such effects. Measuring them by looking for
correlations between Hubble diagram residuals and intrinsic properties
such as delay time, progenitor metallicity, etc. could help correct for them 
(at least statistically if not event-by-event) in future surveys. 

The brightest supernova events only occur in actively star-forming
galaxies (Hamuy et al. 1995, 1996a,b), suggesting prompt explosions, while under-luminous events
are most often found in spirals and E/S0 galaxies,  whose old stellar populations
would suggest delayed explosions \citep{how01}. Mannucci et al. (2006) and \cite{sb05} have
proposed a two-component model for SNe Ia, and several authors \citep{sul06, man05} have
shown that the supernova rate can be expressed as a sum of a term
proportional to the total mass of the galaxy and a term proportional
to the recent star formation rate. In the Mannucci et al. model, some of the supernovae would
explode several Gyr after the birth of the progenitor system, while
others would after a fraction of a
Gyr. Moreover, those two
populations have different luminosities, the ``prompt'' component
being brighter with broader light-curves (Sullivan et al. 2006).
The prompt component will dominate at higher
redshifts when the Universe's age (or the time since star formation
began) was less than the lifetime of the longer-lived progenitors.
Determining
the relative numbers of supernovae in the two populations is the first
step in understanding any possible bias these populations might
cause.
Given the 13\% scatter in the calibrated peak luminosities of
supernovae, measuring systematic effects in the Phillips relation at
the percent level will take samples of a few hundred supernovae \citep{sarkar08}.

Hamuy et al. (2000) used 62 SN Ia host galaxies to study the impact of
host morphology, magnitude and colors on the decline rate $\Delta
m_{15}$, which allows one to estimate the SN peak luminosity.
They found a correlation with age but not with metallicity (see also the erratum to their paper). 
However, their sample was very small and most of their estimates
of age and metallicity were based on photometry only,
without spectra
of the host galaxies, and therefore their accuracy was limited. Moreover, although they
investigated various environmental effects, their
methodology was not sensitive to a second parameter in the
Phillips relation, since they used the decline rate as a ``reddening-free
and distance-free estimate of the SN peak brightness'', and thus
assumed {\em a priori} the universality of the relation.

Gallagher et al. (2005) carried out a similar analysis with spectra of
57 SN Ia hosts, and put tentative constraints
on the SN progenitor lifetime using an estimate of current-to-average star
formation rate.  They claimed to see hints of both a bimodal behavior
and a lower limit of the progenitor lifetime. They admitted that their findings were rather inconclusive.

\citet{sul06} used 100 SNe from the SNLS and broadband spectral energy
distributions of the host galaxies to estimate stellar masses and star formation rates.  They found a component proportional to the stellar mass, and a component proportional to the recent star formation rate, averaged over the last 0.5 Gyr.

Our study improves on these earlier papers by using a larger sample of
SNe, with spectra of their host galaxies from the Sloan Digital Sky
Survey (SDSS; York et al. 2000).  We use a sophisticated stellar
population code called VESPA \citep{vespa} which allows us to
determine the stellar formation history of the hosts.  We also determine
the star formation history of a large sample of normal galaxies from
the SDSS as a control. Differences in those star formation histories will yield
information on SN progenitor lifetime: short lifetimes for instance should
statistically enrich the host sample in galaxies with recent star formation.

\section{Host galaxies and reference sample}

We gathered a sample of about 1300 confirmed SNe Ia from IAU
circulars\footnote{\tt http://www.cfa.harvard.edu/iau/cbat.html}, the CfA
supernovae
list\footnote{\tt http://cfa-www.harvard.edu/iau/lists/Supernovae.html}
and the SDSS-SN public list of
supernovae\footnote{\tt http://sdssdp47.fnal.gov/sdsssn/sdsssn.html}. We
cross-referenced this list with the SDSS DR5 \citep{AM07}
spectroscopic survey of galaxies: 256 galaxies with spectra were
identified as SN Ia hosts, corresponding to 257 supernovae (one galaxy
hosted two supernovae).  The list of hosts used in this paper is
available online\footnote{\tt http://sn.aubourg.net/hosts/}. A large fraction of 
our sample comes from surveys like SDSS (104 SNe), LOSS and LOTOSS (49
SNe for the two surveys). The rest have various origins (Puckett\footnote{\tt http://www.cometwatch.com} : 17
SNe, Pollas (1994) : 12 SNe, etc.).

The detection efficiency of this sample is unknown, as it depends in
detail on the way in which the SNe were found.
To account for the selection function of SN Ia discovery, we also
process a control sample of $10^5$ DR5 galaxy spectra, chosen randomly
in the survey release, and weighted to
reproduce the redshift distribution of the host sample --- this is the
parameter which could most significantly bias delay time measurements. We plan to handle these effects more accurately with a full Monte Carlo treatment in a future paper.

As a cross check, we did the same analysis, keeping only supernovae at $z<0.1$. The reduction in the sample size (from 257 to 190) increases the error, but does not change the result (see table 2 below).
Other effects will be discussed in section 4.

\begin{figure}
\includegraphics[width=\columnwidth,height=8cm]{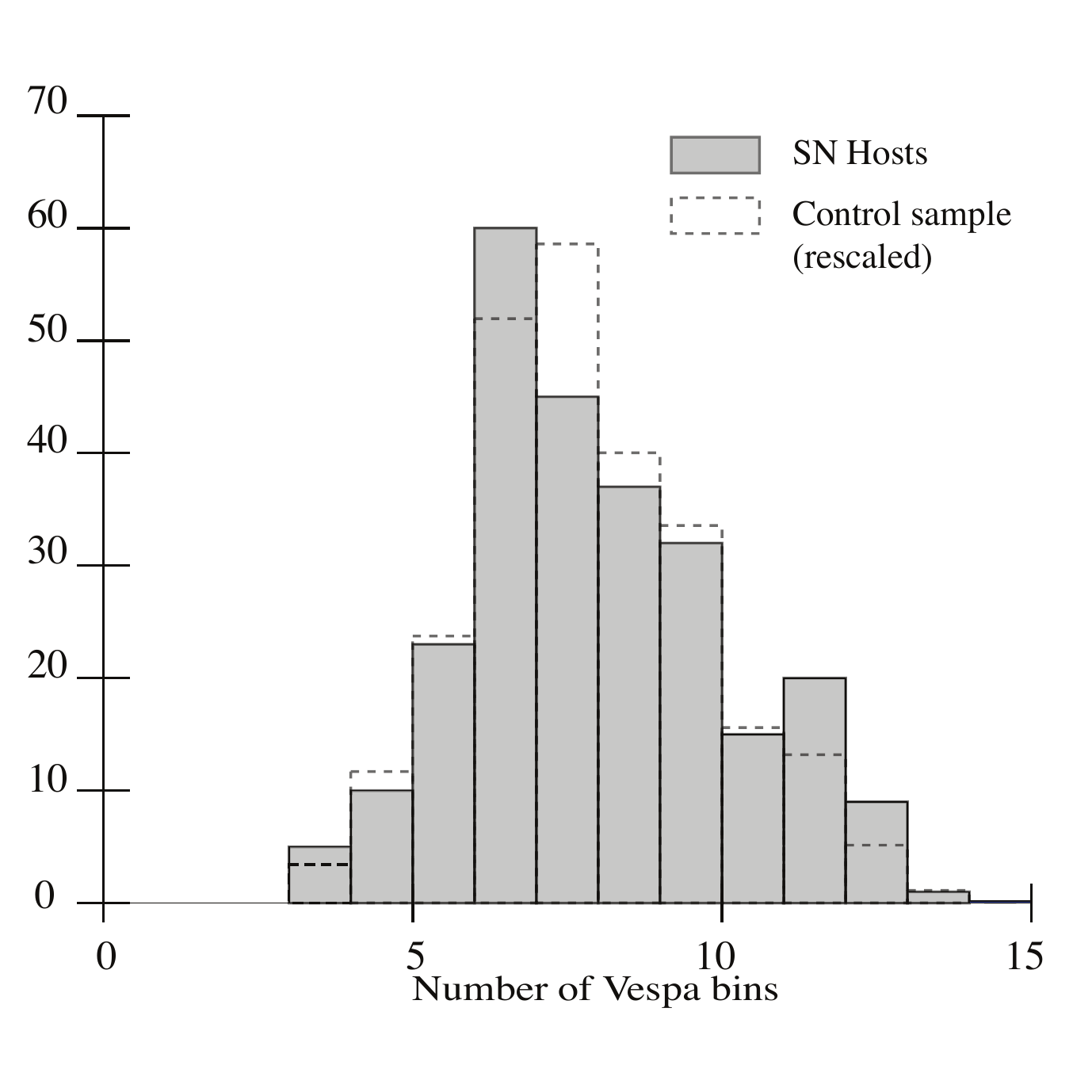}
\caption{Distribution of the number of stellar populations recovered by VESPA from
  the host sample (gray) and the control sample (white). The two
  distributions are very similar (see text).}
\label{fig:vespabins}
\end{figure}

\section{Reconstructing the star formation and metallicity history of SN host galaxies}

The spectrum of a galaxy is a superposition of spectra of single
stellar populations which formed
at a given age with a given metallicity. Since it is not possible to
recover the star formation and metallicity history with infinite
precision \citep[e.g.][]{Jimenez+04}, it is only sensible to attempt
to recover the star formation and metallicity history with a certain
time resolution.  The VESPA algorithm \citep{vespa} does this, providing a
detailed history only where the data warrant it.  Note that broad-band colors are not
sufficient to determine the star formation histories of galaxies, as
they suffer from significant age-metallicity degeneracies
\citep[e.g.][]{Jimenez+04}.

\begin{figure*}
\includegraphics[width=2\columnwidth]{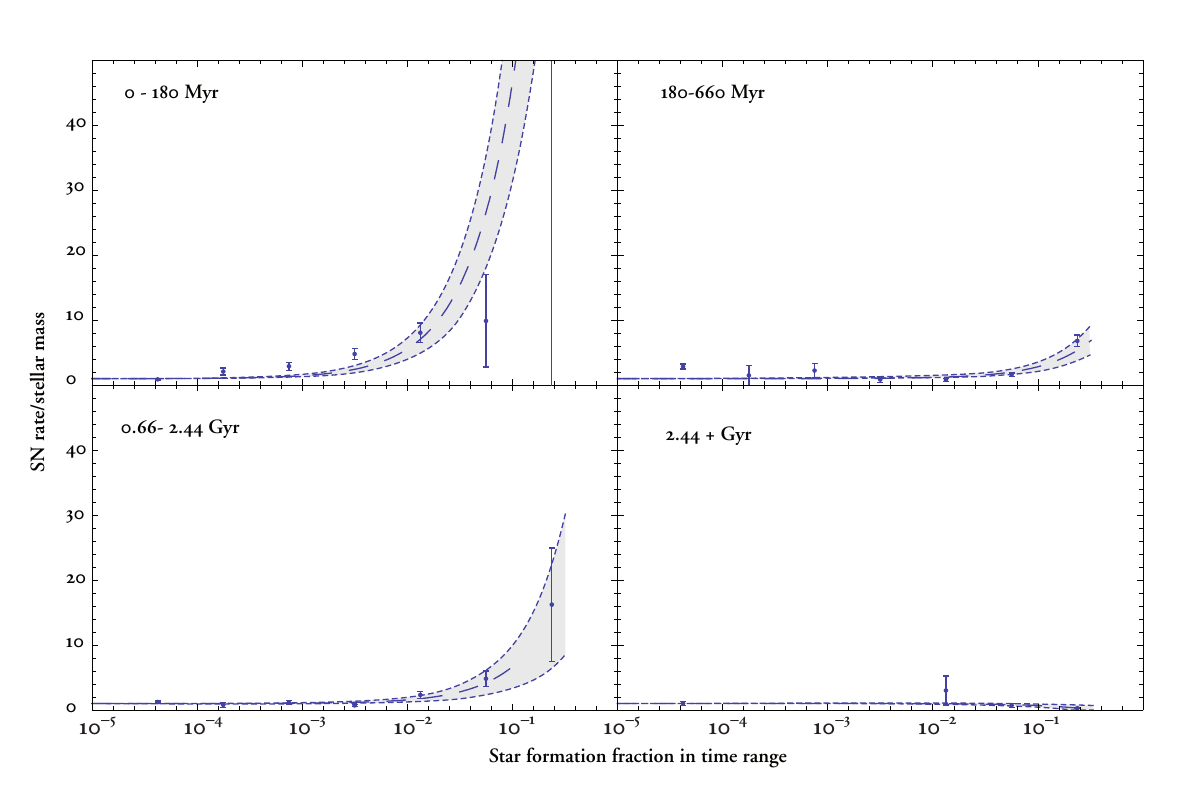}
\caption{Type Ia supernova rate per stellar mass, unnormalized,
  vs. fraction of stellar mass formed in the four time intervals
  indicated in each panel.  The dashed line is a fit to a dual
  component model ${SNR} = \alpha \times \mstar + \beta \times  M_{\rm
    time\ range}$, for each time range.  In the absence of a stellar
  population of a given age contributing to the supernova rate, the
  curve would be consistent with flat (i.e,, $\beta$ consistent with
  zero). We find that
  $\beta/\alpha$ is non zero at the five-sigma level for the time
  intervals [0-180] Myr. Shaded areas are
  $\pm2\sigma$ fit values.} 
\label{fig:rate}
\end{figure*}

In brief, VESPA uses singular value decomposition to calculate the
number of significant components in the spectrum of a given galaxy. VESPA
then uses an algorithm to determine the best-fitting
non-negative values of the star formation fractions.    Extensive
tests of the performance of VESPA on synthetic spectra as a function
of wavelength coverage and signal-to-noise ratio can be found in
\citet{vespa}, along with a study of age-metallicity degeneracy.
To limit the search to a
manageable amount of parameters, and because currently available
spectra never have the quality or spectral range to justify going
beyond this choice, VESPA's finest resolution consists of 16 age bins,
logarithmically spaced in lookback time between 0.002 and 14 Gyrs.
Specifically, the lower limit in age of
the 16 bins are: 0.002, 0.02, 0.03, 0.0462, 0.074, 0.115, 0.177,
0.275, 0.425, 0.6347, 1.02, 1.57, 2.44, 3.78, 5.84, and 9.04 Gyr.

VESPA chooses the number of stellar populations to model depending on
the quality of the data.  The SDSS galaxy spectra typically allow
between 7 and 10 age bins in both the SN host
sample and the control sample,  showing there are no significant differences in the two samples
 (Fig.~\ref{fig:vespabins}; see also
Tojeiro et al. 2007), although
there is non-zero star formation in only 3-5 of those bins.
The relatively large number of bins implied by Fig.~\ref{fig:vespabins} is an indication that in most cases the star formation which is recovered is localised to narrow time intervals. 
The metallicity for each population is a free parameter, so
there are as many metallicity values recovered as there are star
formation fractions.  

Following Sullivan et al. (2006), we parametrize the total supernova
rate with a two-component model, ${SNR} = \alpha \mstar + \beta
\mrecent$, where $\mstar$ is the total stellar mass and $\mrecent$ is the
mass formed in some time range that we can vary. $\alpha$ and $\beta$ reflect,
respectively, the SNR per unit stellar mass of an old population of
progenitors (roughly proportional to the total stellar mass), and the SNR per
unit stellar mass of a young population of progenitors (proportional to
the mass in recently formed stars). Because we do
not have a calibration of the selection function of the supernova
surveys we have used, we normalize with our control sample, representative of the whole SDSS spectroscopic catalog (our reference sample), and can
determine $SNR$ only up to a proportionality constant.  That is, we
fit the data to a model 
\begin{equation} 
\frac{SNR}{\mstar} = C\left(1 + \frac{\beta}{\alpha}
\frac{\mrecent}{\mstar}\right),
\end{equation}
where $C$ is an unnormalized proportionality constant.  
Thus the quantity $\beta/\alpha$ quantifies the fraction of the
supernova rate with progenitor stars that formed in the chosen time range. 
The constant $C$ is the product of the ``slow'' rate $A$ introduced in
the Mannucci et al. two-component model, and an unknown global
SN detection efficiency factor averaged over all SN surveys which
contribute to our reference sample.

We estimate the SNR per unit stellar mass, $SNR/\mstar$ as a function
of $\frac{\mrecent}{\mstar}$ as follows.  For any unbiased sample of
galaxies, the observed SNR per unit stellar mass is equal to the
number of supernovae divided by the total stellar mass in the sample,
multiplied by an efficiency factor $\epsilon$: $\frac{SNR}{\mstar} =
\epsilon \frac{N_{SN}}{M_{total}}$. We can divide them
into bins of their star formation fraction in the recent time range,
$\frac{\mrecent}{\mstar}$, as determined by the VESPA analysis. Within
each bin we can therefore calculate $\frac{SNR}{\mstar}$, by simply
counting the number of SNe in each subset and dividing it by the total
stellar mass in galaxies with this value of
$\frac{\mrecent}{\mstar}$. The latter is calculated from the much larger
control sample. 

We repeated this exercise for a variety of ages for the young component, and
also varying the dust model details in VESPA and the bin boundaries.
We find a significant contribution to the SN rate from recent stellar
populations (in the sense that $\beta/\alpha$ is significantly
positive), where we vary the definition of ``recent'' between 74 and
180 Myr.  
The contribution of stars in the 180-250 Myr range is
lower by at least a factor of five. 
The value of $\beta/\alpha$ is robust to setting the boundary anywhere
between 74 and 180 Myr, and thus here we quote the most conservative
value of 180 Myr, since starburst duration and internal degeneracies in VESPA could
make age bins ``leak'' one into another.

We illustrate this in Figure~\ref{fig:rate}.  We have divided our 16
time bins into four broader bins, and asked for the correlation of the
supernova rate with the fraction of star formation that occurred in each bin.  The
four panels show the un-normalized Type Ia supernova rate per stellar
mass as a function of the fraction of stellar mass formed each of
these broad time bins [0-180] Myr (top-left panel), [180-660]Myr
(bottom-left panel), [0.66-2.44] Gyr (top-right panel), and [2.44-13]
Gyr (bottom-right panel). The resulting best-fit value of
$\beta/\alpha$ and the corresponding correlations are given in Table
1. 

For the youngest range, 0-180 Myr, we measure $\beta/\alpha = 454 \pm 78$, which is five-sigma evidence for a short
duration component. 
Since $\beta/\alpha$  in older bins is significantly lower, the
positive signal we find in the most recent age range cannot be due to
correlation (``leakage'') from older bins.

\begin{table}
\begin{tabular}{|c|c|c|c|}
\hline
Age (Myr) & $\beta/\alpha$ & Error on $\beta/\alpha $& Significance ($\sigma$)\\
\hline
0-180 & 454 & 78 & 5.8 \\
180-660 & 56 & 16 & 3.4 \\
660-2440 & 18.4 & 3.5 & 5.2 \\
2440-13700 & -3 & 1.0 & -- \\
\hline
\end{tabular}
\caption{Fits of supernova rate per unit stellar mass $\propto 1 + 
  + {\beta\over\alpha} \frac{M_{\rm range}}{M_*}$ for different ranges of star
  formation lookback time for the $\beta$ component.}  
\end{table}

\begin{table}
\begin{tabular}{|c|c|c|c|}
\hline
Age (Myr) & $\beta/\alpha$ & Error on $\beta/\alpha $& Significance ($\sigma$)\\
\hline
0-180 & 347 & 75 & 4.6 \\
180-660 & 43 & 20 & 2.2 \\
660-2440 & 46 & 8 & 5.8 \\
2440-13700 & -3 & 1.0 & -- \\
\hline
\end{tabular}
\caption{Same as table 1, considering only SNe with $z<0.1$.}  
\end{table}

Our results should be robust against possible spectroscopic calibration
errors: because we compare
the host population to a control sample of spectra taken with the same
telescope and instrument, and processed with the same
spectroscopic pipeline and the same star formation history recovery
algorithm (VESPA), systematic errors anywhere in the chain would be shared by the
two samples. 

The selection function of our SN sample is also irrelevant as long as it
is not strongly dependent on host properties which are correlated with SFR;
the effect of possible efficiency biases in the SN host sample will be
discussed in more detail in \S 4 below.  
We do not consider supernovae without a SDSS host but this
is fully consistent with our approach of comparing to a SDSS
reference population : our reference sample is, by definition, complete, and we compare, among that sample, hosts without SN to hosts with SN.

A bias will be introduced in our results if there is a correlation
between galaxy mass and specific star formation rate.  Indeed, most
massive galaxies in the present-day universe have little star
formation; most of the star formation in the present-day universe is
in low-mass galaxies (Heavens et al. 2004).  We have tested this bias with
a Monte-Carlo simulation, as follows. 

\begin{figure}
\includegraphics[width=\columnwidth]{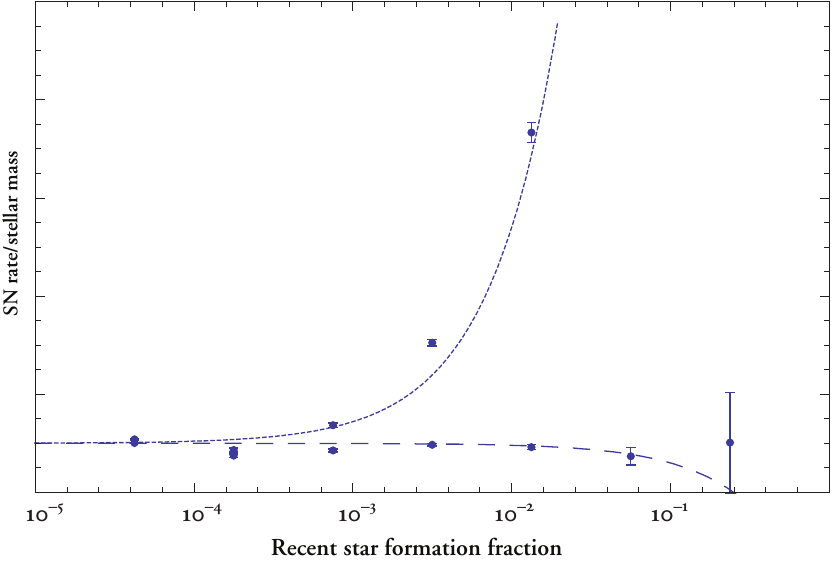}
\caption{Estimation of the possible bias that can be introduced by star formation today taking place preferentially in smaller mass galaxies. Using our SDSS control sample we have generated supernova using $\beta/\alpha = 0$ and $\beta/\alpha = 700$. The recovered lines show that for the case with $\beta/\alpha = 0$ (dotted line) we indeed find no signature of supernova, while for the case $\beta/\alpha = 700$ we do recover supernova activity at the $2/3$ level (dotted line) of the input value ($\beta/\alpha \simeq 465$), thus our conclusions are conservative. See more details in text.}
\label{fig:bias}
\end{figure}

We generated an artificial SN sample in our control sample following
the rule 
$ \alpha \times \mstar + \beta \times  M_{\rm recent}$.  In the absence of correlations
between recent star formation and mass in this reference sample, applying our method
recovers the input $\beta/\alpha$ values.  However, our reference sample has
an anti-correlation between mass and recent star formation.
With the
input $\beta$ set equal to zero, we recover $\beta/\alpha = -3 \pm 2$,
compatible with zero (dashed line in
Fig.~\ref{fig:bias}). Using $\beta/\alpha = 700$ (dotted line in Fig.~\ref{fig:bias}), we 
recover $\beta/\alpha = 465$, $2/3$ of the simulated value (and close
to our observed value).  
 Therefore we conclude that the anti-correlation between galaxy mass and star
 formation rate causes us to {\em underestimate} the contribution
 of a prompt component to the supernova rate.

The ratio $\beta/\alpha$ is compatible with previous estimates, for
which the ``recent'' SFR is estimated in general from colors,
broadband SED fitting, core-collapse SN rate or cosmic SFR.  These
results represent an
average over half a gigayear, and are thus only a rough match to our
results.
We can roughly convert our mass estimate  $M_{180}$ ($M_{recent}$ for the last 180 Myr)
 to recent SFR through $M_{180} =
180\times 10^{6} \frac{SFR}{1 M_\odot \rm yr^{-1}} M_\odot$.  With this, the Neill et
al. (2006) values (a ``slow'' rate $A = 1.4 \pm 1.0 \times 10^{-14}$ SN
$M_\odot^{-1}$ $\rm yr^{-1}$ and a ``prompt'' rate $B = 8.0 \pm 2.6 \times
10^{-4}$ SN $(M_\odot \rm yr^{-1})^{-1} yr^{-1}$) yield
$\beta/\alpha \simeq 300$, the Sullivan et al. (2006) values 
($5.3 \pm 1.0 \times 10^{-14}$ SN
$M_\odot^{-1}$ $\rm yr^{-1}$ and $3.9 \times
10^{-4}$ SN $(M_\odot \rm yr^{-1})^{-1} yr^{-1}$) yield
$\beta/\alpha \simeq 40$, and the two values quoted by Scannapieco \&
Bildsten (2005) yield $\beta/\alpha \simeq 300$ and 150,
respectively. However since ``recent'' star formation is estimated differently in each case, and on an difference recent time range, the figures are not expected to match closely.

\section{Discussion and conclusion}
Our result would be sensitive to any systematic effect enriching the
SN host sample in blue galaxies (i.e., those with large $M_{recent}/\mstar$),
for reasons unrelated to SN physics (bias in efficiency, or bias in
the monitored galaxy sample for targeted searches).
Some targeted supernova searches have focused on star-forming galaxies (e.g.,
Richmond, Filippenko, \& Galisky 1998).  
In our sample, the main targeted search is the Lick
Observatory Supernova Search (LOSS, Filippenko et al. 2001). There is no indication this
search is biased in this way, and our results change insignificantly if we remove those hosts.

Blue galaxies are fainter, and one could naively expect to detect SNe more
easily in those faint hosts, but with modern image differencing
techniques, the effect should be minor. One also expects such a bias to occur in
surveys that rely on spectroscopic typing of SNe, since the fainter, low-stretch SN, that occur preferencially in bright galaxies, will be more difficult to characterize spectroscopically.
Star-forming galaxies tend to produce brighter events.  On the other
hand, star-forming hosts tend to be dustier, making supernovae harder
to detect or characterize.  If such an efficiency effect had an impact on our
analysis, in the sense of enhancing or simulating what we observe, we
would expect the host galaxy sample to exhibit lower typical masses
than the reference sample, or lower dust content.  VESPA yields an
estimate of dust content and luminosity of the hosts. We see no
significant difference in luminosity and dust distributions between
the host and control samples.  Rather, the host stellar mass
distribution is shifted towards slightly {\em higher} masses, as we
would expect given that the SN rate should increase with the mass. We
are thus confident this effect does not significantly bias our
results.

Such an efficiency effect would also be more important at high redshift. As we noted above,
restricting our sample to SNe occurring at $z<0.1$ does not change our findings. In our next
paper, we plan to use the lightcurve characteristics in our analyses
for those objects for which they are available.

Following Mannucci et al. (2005, 2006) and Sullivan et al. (2006), we have shown
that SNe Ia can occur through short-lived progenitors, hinting at a
variety of stellar evolution paths with different lifetimes. We have
given the first estimate of the lifetime of the ``fast''
component, by reconstructing the star formation history of SN host
galaxies and finding an increased contribution to the SN Ia rate from
stars evolving in less than 180 Myr.

Such a short time delay strongly constrains the nature of possible
progenitors. They must be stars that evolve fast enough, i.e. with a
mass above  $\sim 3.5 M_\odot$, but must be below the super-AGB mass threshold (about 8
$M_\odot$) above which one gets electron-capture supernovae
\citep{poel07}. \citet{PS06} have
also suggested that a significant fraction of
binaries are twins (i.e., pairs of stars with essentially identical
masses), and that such twin binaries could produce a short
($< 0.1$ Gyr) path to SN Ia.  Considering
common envelope evolution phenomena, \citet{PS06} argue that such twin systems
could yield double degenerate SNe Ia in a way that would be both fast
and efficient (see also \citealt{hachisu}).

Are there enough high-mass progenitors to account for
the observed SN Ia rate? These progenitors have to have
masses between $3.5 M_\odot$ (in order to explode within 180 Myr)
and $8 M_\odot$. Only
a fraction of these stars $f_\beta$ will actually
explode as a SN Ia progenitor. We take into account five factors: the fraction of stars in binaries ($f_a$), the fraction
of the binaries both of whose components lie in the range 3.5 to 8
$M_\odot$ ($f_b$), the fraction
of stars at a suitable separation for mass transfer ($f_c$), the
fact that every binary yields
a single explosion, ($f_d$), and an overall efficiency ($\eta_\beta$,
as not all possible progenitors may explode). Maoz (2007) has estimated
the first four factors, and finds $f_a \in [2/3,1], f_b\in[1/6, 1/3],
f_c\in[1/4,1/2]$ and $f_d=1/2$. Crudely multiplying these factors together
gives the fraction of objects in the appropriate mass range that
explode as prompt SNe Ia: $f_\beta\in[0.014, 0.083]\eta_\beta$.

The fraction of stars which will explode as a SN Ia progenitor,
$f_\beta$ is then given by $f_\beta = \frac{N_B}{N_{3.5-8}}$ where $N_B$ is
the total number of SNe Ia from the fast route, and $N_{3.5-8}$ is the
total number of stars in the correct mass range. Our result only
allows us to estimate the ratio of the two components.   We can however
use published values of $A$ in order to infer an absolute value of $B$
from our ratio---we call this value $B'$. Using $A=1.4 \pm 1.0 \times10^{-14}
SN yr^{-1}M_\odot^{-1}$ as published by Neill et al. (2006) gives $B' =
6.3 \pm 4.7 \times 10^{-12} SN\rm \, yr^{-1}\, M_\odot^{-1}$. We can now
estimate $N_B
= B' \times 180\,{\rm  Myrs} \times M_{180}$ and $N_{3.5-8} =
0.0157*M_{180}$, assuming a Salpeter IMF of the form $N(m) \propto
m^{-2.35}$. This gives $f_\beta = 0.073 \pm 0.053$, within the values
given by Maoz (2007). Using a value of $A=5.3 \pm 1.1 \times 10^{-14}$ from
Sullivan et al. (2006) gives a value of $B' = 2.41 \pm 0.65 \times 10^{-11}$
and $f_\beta = 0.28 \pm 0.07$. Similarly, $A=4.4^{+1.6}_{-1.4}
\times 10^{-14}$ as published
by Scannapieco \& Bildsten (2005) results in roughly $B' = 1.99 \pm
0.69 \times 10^{-11}$ and $f_\beta = 0.23 \pm 0.08$. 

With the slow $A$ rate of Neill et al (2006) there is complete
consistency with the theoretical expectations of SN Ia rates from Maoz
(2007).  However, with the
higher $A$ rates of Sullivan et al (2006) or Scannapieco \& Bildsten (2005), there is some
tension with our results, which would require high efficiency. It is
hard to know how worried one should be about this: firstly we are
relying on external measurements of the delayed rate, and secondly an
excess of SNIa explosions with respect to the predicted number of
progenitors is observed in a large number of SNIa studies (Maoz 2007).
This tension can be alleviated in a number of ways---some of which
discussed in the above paper---but generally indicates that the
efficiency of the mechanism which produces SNIa explosions must be
high.

A crucial question for the use of SNe Ia as standard candles
in cosmology is whether these different routes yield objects which are
standardizeable to high accuracy via the same empirical corrections.
Current data find
no evidence for a difference \citep{hamuy00,sul03,bronder07},
but the requirements for using SNe Ia as
a Dark Energy probe are stringent, and it will be important to
establish this point accurately.  VESPA should be able to assist
directly with the correction, first by determining
the standardization for each route at the required
percent level accuracy, and then allowing to apply the correct
standardization at least statistically if not
for individual supernovae. 
  Furthermore, VESPA provides metallicity estimates for the hosts,
  which may correlate with peak luminosity, and thus allow us to
  reduce the scatter in the distance indicator.  This will be the
  subject of a future study. 

We plan to expand our sample by obtaining spectra of a larger number
of SN hosts, allowing us to deconvolve the delay time
function. Future papers will address more quantitatively the long
duration component, the downsizing bias, the metallicity effects (see also Prieto et
al. 2007), and stellar evolution models compatible with our
findings.

EA acknowledges the importance of numerous discussions with the late
Bohdan Paczy\'nski, to whom we dedicate this paper. RT is funded by the Funda\c{c}\~{a}o para a Ci\^{e}ncia e a Tecnologia under the reference PRAXIS SFRH/BD/16973/04. RJ and DNS
acknowledge support from the NSF PIRE-0530095. This work was supported
in part by  DOE grant DE-FG02-07ER41514.

\end{document}